\documentclass[twocolumn]{aastex631}

\newcommand{\spi}{{Spitzer}}
\newcommand{\wise}{{WISE}}
\newcommand{\iras}{{IRAS}}
\newcommand{\iso}{{ISO}}
\providecommand{\mum}{\mbox{$\mu$m}}

\providecommand{\msun}{M$_{\odot}$}
\providecommand{\water}{\mbox{{H$_2$O}}}

\submitjournal{ApJ}\received{2024 March 4}
\revised{2024 July 26}
\accepted{2024 August 8}
\published{2024 September 30}

\shorttitle{SOFIA FORCAST S Stars}
\shortauthors{Kraemer et al.}

\begin{document}

\title{The Dustiest Galactic S Stars: Mid-Infrared Spectra from SOFIA/FORCAST}

\correspondingauthor{Kathleen E. Kraemer}
\email{kathleen.kraemer@bc.edu}

\author[0000-0002-2626-7155]{Kathleen E. Kraemer}
\affiliation{Institute for Scientific Research, Boston College,\\
140 Commonwealth Avenue, Chestnut Hill, MA 02467, USA}

\author[0000-0003-4520-1044]{G. C. Sloan}
\affiliation{Space Telescope Science Institute, 3700 San Martin
Drive, Baltimore, MD 21218, USA}
\affiliation{Department of Physics and Astronomy, University of
  North Carolina, Chapel Hill, NC 27599-3255, USA}

\author[0000-0001-7553-8444]{Ramses M. Ramirez}
\affiliation{University of Central Florida, Department of Physics, \\
Planetary Sciences Group, Orlando, FL 32816, USA}

\begin{abstract}
We present spectra of 12 of the reddest, and hence dustiest, S stars in 
the Milky Way, observed with the FORCAST grisms on SOFIA. S stars are 
asymptotic giant 
branch (AGB) stars with C/O$\sim$1, so their molecular and dust chemistries are 
dominated by neither O nor C, often leading to atypical spectral features from 
their molecules and dust grains. All of the stars in our sample have strong 
dust emission features at 10--11 \mum, but the shape of the feature in most 
of the stars differs from the shapes commonly observed in either 
oxygen-rich or carbon-rich AGB stars. Two stars also show the 13 \mum\ feature 
associated with crystalline alumina. Two have a water absorption
band at $\sim$6.5--7.5 \mum, and a third has a tentative detection, but 
only one of these three has the more common SiO absorption band at 7.5 \mum. 
Three others 
show a red 6.3 \mum\ emission feature from complex hydrocarbons consistent
with ``Class C'' objects, and in a fourth it appears at 6.37 \mum, redder than 
even the standard Class C hydrocarbon feature. Class C spectra 
typically indicate complex hydrocarbons which have been less processed 
by UV radiation, resulting in more aliphatic bonds relative to aromatic bonds. 
None of the S stars shows a strong 11.3 \mum\ hydrocarbon feature, which is 
also consistent with the presence of aliphatic hydrocarbons. 

\end{abstract}


\section{Introduction} \label{sec.intro}

A low- to intermediate-mass star, M$\sim$0.8--8 \msun,
ejects the bulk of its mass while on the asymptotic giant branch (AGB), 
ultimately ending its life as a white dwarf with M$<$1.4 \msun. This ejected
material will be highly enriched with fresh fusion products, contributing
significantly to the chemical enrichment of the Galaxy and the Universe
\cite[e.g.,][]{burbidge+57, ulrich73,kl14,kobayashi+20}. 
As the ejecta cool, molecules and grains condense, forming a circumstellar
envelope that has a rich spectrum of features in the infrared, from which
the chemical and physical conditions of the material can be inferred 
\cite[e.g.,][and references therein]{hab96,ho18}.

CO is the molecule which determines the chemistry in the outer envelopes of
cool AGB stars because it consumes most of the available C or O, whichever
is less abundant.  On the AGB, freshly fused carbon is dredged up from the
interior, and if the C/O ratio exceeds 1, the chemistry of the gas
and dust will change dramatically to reflect that new balance 
\cite[e.g.,][]{iben74, ir83, wk98}.
Although there is a continuum of C/O ratios in AGB stars,
the chemistry of the gas and dust shows a dichotomy. Carbon-rich chemistry
dominates in carbon stars and oxygen-rich chemistry
dominates most of the remaining AGB stars (M giants), and the mid-infrared
spectra of each type reflect this dichotomy with distinct sets of
molecular absorption and dust emission features 
\cite[e.g.,][]{hackwell72, merrillstein76, cheesemanea89,kspw02,matsuuraea05,ruffleea15,jonesea17}.

S stars lie in the transition between these two chemical 
domains, with C/O$\sim$1. Because of this, odd chemistry can occur in
the stellar atmosphere and circumstellar envelope, as different chemical paths
are available due to the lack of both C and O, one of which would normally 
dominate the chemistry. S stars and their infrared spectra can thus 
directly probe a chemical regime of circumstellar material that is not
present in studies of AGB stars which focus on either the carbon stars or the
oxygen-rich M giants. 

While the infrared spectra of these stars are commonly classified as part of the
oxygen-rich population \cite[e.g.,][]{sp98,kspw02}, 
their spectral features are often not those typical of silicate and alumina 
dust and the associated molecular features
\cite[e.g.,][]{lml88,honyea09, smoldersea12a}. For example, the
shape of the 10 \mum\ feature can shift to 10.5 \mum,
and many lack the 18 \mum\ silicate
emission feature altogether. Others show features more typically seen in
carbon stars such as complex hydrocarbons (e.g., polycyclic aromatic
hydrocarbons or PAHs) and weak versions of
the 26--30 \mum\ feature attributed to MgS \cite[e.g.,][]{honyea09,
smoldersea10, smoldersea12a}. Gaseous SiS bands may be seen at 6.7 and 
13--14 \mum\ in absorption \citep{aokiea98, camiea09,smoldersea12a}, 
and in rare cases in emission \citep{shs11}.

The samples in those previous spectroscopic studies, however, have generally 
undersampled the dustiest, i.e., reddest, S stars. We therefore undertook a
project using the Faint Object infraRed CAmera for the SOFIA Telescope 
\cite[FORCAST,][]{forcast} on the Stratospheric Observatory for Infrared 
Astronomy \cite[SOFIA,][]{sofia} to address this gap.
Section \ref{sec.sources} describes our source selection, the observations, 
and the data processing. Section 3 describes the results and analysis, and
Section 4 summarizes
our findings.

\section{Source Selection and Observations} \label{sec.sources}
\begin{figure}
\includegraphics[width=3.5in]{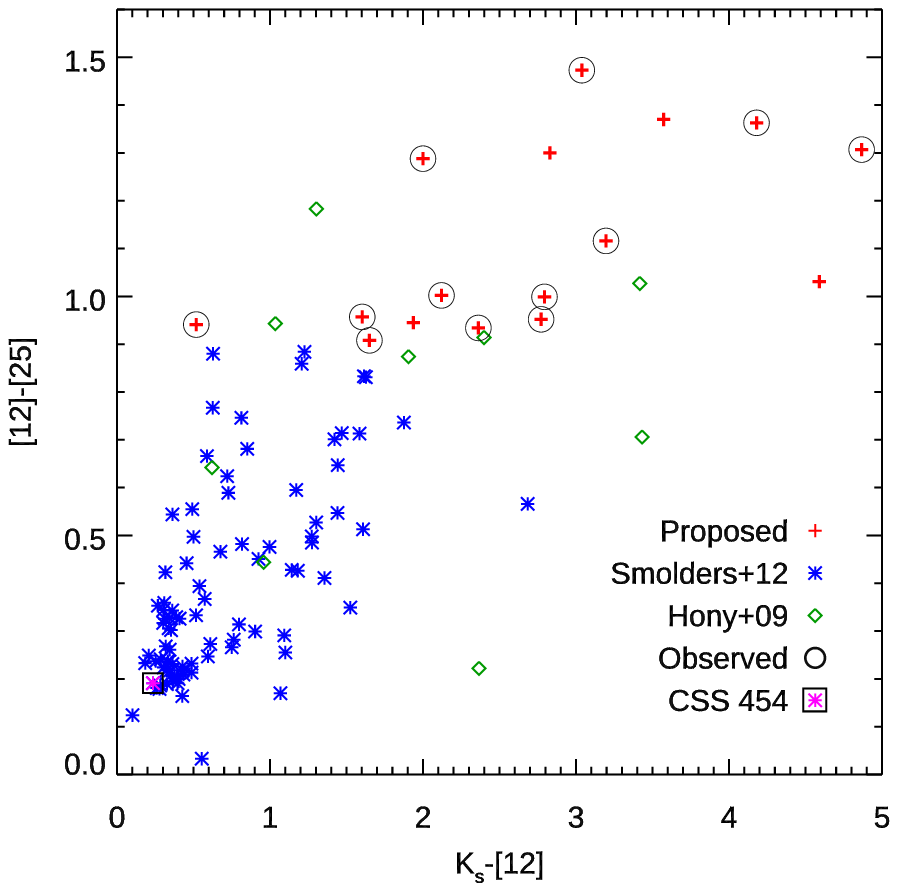}
\caption{IR color-color diagram for the S star samples with mid-IR spectra. 
\cite{honyea09}: green diamonds; \cite{smoldersea12a}: blue asterisks; our
proposed sample: red pluses; those observed with FORCAST: circles; 
magenta asterisk with black box: CSS 454, the star used for 
continuum subtraction in Sec. \ref{sec.dust}.
}
\label{fig.colcol}
\end{figure}
Our goal was to investigate the dustiest S stars as 
these were underrepresented in previous studies. We started with the S stars
listed in Simbad\footnote{As of July 2020} and checked the database of 
\cite{chenea19} to ensure we selected intrinsic S stars, i.e., those 
whose chemical balance is due to the dredge-up of carbon from the 
interior, rather than contamination from mass transfer in a binary. 
To select the dustiest stars, we used  12 and 
25 \micron\ photometry from the Point Source Catalog (PSC) of the Infrared 
Astronomical Satellite \cite[\iras][]{iras84}, complemented by  K$_s$ photometry from the 
2 Micron All-Sky Survey \cite[2MASS,][]{iras84}. We required  
[12]$-$[25] $>$ 0.9 as these would be the 
dustiest stars and are the most under-sampled, as Figure \ref{fig.colcol} 
shows. Since the  K$_s-$[12] color is not
as sensitive to the coolest dust as the mid-infrared color, it was used as the
secondary constraint. In particular, we
preferentially chose stars with  K$_s-$[12] $>$ 1.0; only one of our targets
is bluer than this. The \iras-based colors were compared with those from  the
{\em Wide-field Infrared Space Experiment} \cite[\wise;][]{wise10}, and
most (13 of 16) also had $W3-W4>0.9$. \wise-based colors were not used as 
the primary criterion as these sources are almost all partially saturated in 
the \wise\ bands\footnote{Only W4 for HD 35273 was unsaturated.}. 
With one 
exception, sources which
had been observed with the Short-Wavelength Spectrometer \cite[SWS,][]{sws96}
on the Infrared Space Observatory \cite[ISO,][]{iso96}
were removed from the candidate target list. RW And was retained as a target since 
its SWS spectrum was too noisy to use \citep{kspw02, honyea09}.
A lower 
limit to the 12 \micron\ flux density of $F_{12}=10$ Jy
ensured the required signal-to-noise ratio could be reached in reasonable 
integration times. We also imposed a declination limit
of $-$25\arcdeg\ as FORCAST was not expected to deploy to New Zealand in 
Cycle 9. 

These selection criteria resulted in a set of 16 S stars, of which 12 were 
observed with the FORCAST grisms under Plan ID 09\_0046.  Table \ref{tab.obs} 
provides details on the observed targets; those not observed are 
listed in the Appendix. Eight stars have data from all four 
grisms: G063 (4.9--8.0 \mum), G111 (8.4--13.7 \mum), G227 (17.6--27.7 
\mum), and G329 (28.7--37.1 \mum). The other four were observed in at least
the G111 grism. 

The observations used the two-position chop and nod setting (C2N) 
with the
default 60\arcsec\ chop throw and 30\arcdeg\ chop angle. Integration times 
were set to achieve a signal-to-noise
ratio of S/N$>$30 in G063 and G111 and S/N$>$25 for G227 and G329 (the 
features in the redder grisms typically being broader than those in the 
bluer grisms). All times in G063 were set to 30 sec. Those in G111 and 
G227 were set to 30, 60, or 100 sec, depending on the 12 and 25 \mum\ flux 
density. Those
in G329 were set to 100, 300, or 500 sec, depending on the 25 \mum\ flux
density. These settings would achieve our S/N goals based on the SOFIA 
Instrument Time Estimator (SITE).\footnote{No longer
available after mission end. The minimum integration time of 30 sec was set
by the fact that observations this short were dominated by overheads.} 
The G329 grism was omitted for sources with $F_{25}$ $<$15 Jy.
The 4\farcs7 slit was used for all observations, which results in spectral 
resolving powers of R$\sim$125 for G063 and  G111 ($\lambda$=4.9--13.7 \mum)
and R$\sim$70--110 for G227 and G329 ($\lambda$=17.6--37.1 \mum).

We use the Level 3 spectra provided by the SOFIA pipeline. The pipeline
corrects instrumental effects, extracts spectra, applies
the flux calibration, corrects for atmospheric transmission effects, 
and co-adds any spectra taken at the same settings to generate a single 
spectrum for each grism.

Only the G111 spectral segment required additional
processing. For this segment, the atmospheric correction left residuals,
notably in the ozone band at $\lambda$ $\sim$9.25--10.05 \mum. Also,
the reddest data were usually very noisy due to residual \water\ and CO$_2$
vapor long-ward of $\sim$13.8 \mum. These spectral regions were therefore 
masked out individually for each star.

\begin{deluxetable*}{lrrrrrrrrrr}
\tablecaption{Source Properties and Observation Summary\label{tab.obs}}
\tablewidth{0pt}
\tablehead{
\colhead{Star} & \colhead{RA} & \colhead{Dec}  &
\colhead{$F_{12}$} & \colhead{$F_{25}$} &\colhead{[12]$-$[25] }  &
\colhead{K$_s-$[12]}& \colhead{Spectral} & \colhead{Obs. Date} & 
\colhead{$\lambda_{obs}$}\\
\colhead{Name} & \colhead{(J2000)} & \colhead{(J2000)}
& \colhead{(Jy)} & \colhead{(Jy)} & \colhead{(mag)} 
& \colhead{(mag)}& \colhead{Type\tablenotemark{a}}& \colhead{(yyyy mmm dd)}& \colhead{(\mum)} 
}
\startdata
RW And       &  11.828799 &    32.685600  & 36.1 & 18.5 &  0.91 &  1.65 & S6/2e &2022 May 26 & 5.1--37.1\\
HD 35273     &  80.836281 &  $-$4.570621  & 20.0 & 10.7 &  0.96 &  1.60 & M4wkS &2022 Feb 1 & 5.1--13.8 \\
Y Lyn        & 112.048424 &    45.990589  &121.9 & 64.2 &  0.94 &  0.52 & M6S   &2022 Feb 8 & 5.1--37.1\\
IRC $-$10401 & 272.603333 & $-$10.571130  &177.4 &109.8 &  1.12 &  3.20 & M7S   &2022 Feb 2 & 5.1--37.1\\
IRC $-$10411 & 275.099243 & $-$14.113032  & 37.9 & 27.5 &  1.29 &  2.00 & S     &2021 Jul 9 & 5.1--37.1\\
CSS 1055     & 278.184753 &  $-$9.486253  & 28.4 & 15.8 &  1.00 &  2.79 & S     &2021 Jul 8 & 5.1--37.1\\
IRC $-$10450 & 280.056274 &  $-$5.703139  &124.4 & 91.8 &  1.31 &  4.87 & S     &2021 Jul 9 & 5.1--37.1\\
IRC +00402   & 284.601105 &     4.665304  & 88.2 & 46.2 &  0.93 &  2.36 & M2    &2022 May 25 & 5.1--37.1\\
CSS2 41      & 294.782379 &    29.044159  & 50.4 & 39.1 &  1.36 &  4.18 & S     &2022 Feb 10 & 5.1--37.1\\
CSS 1185     & 300.791840 &    29.986435  & 29.7 & 15.8 &  0.95 &  2.77 & S     &2022 May 26 & 5.1--27.9\\
IRC +60374   & 343.301422 &    61.283463  &108.5 & 93.3 &  1.47 &  3.04 & M3Ib  &2022 Sep 14 & 8.8--13.8\\
WY Cas       & 359.505493 &    56.487099  & 50.9 & 28.4 &  1.00 &  2.12 & S6    &2022 Jan 27 & 5.1--27.9\\
\enddata
\tablenotetext{a}{From Simbad}
\end{deluxetable*}

\section{Results and Discussion\label{sec.res}}

\begin{figure*}[ht] 
\includegraphics[width=6.5in]{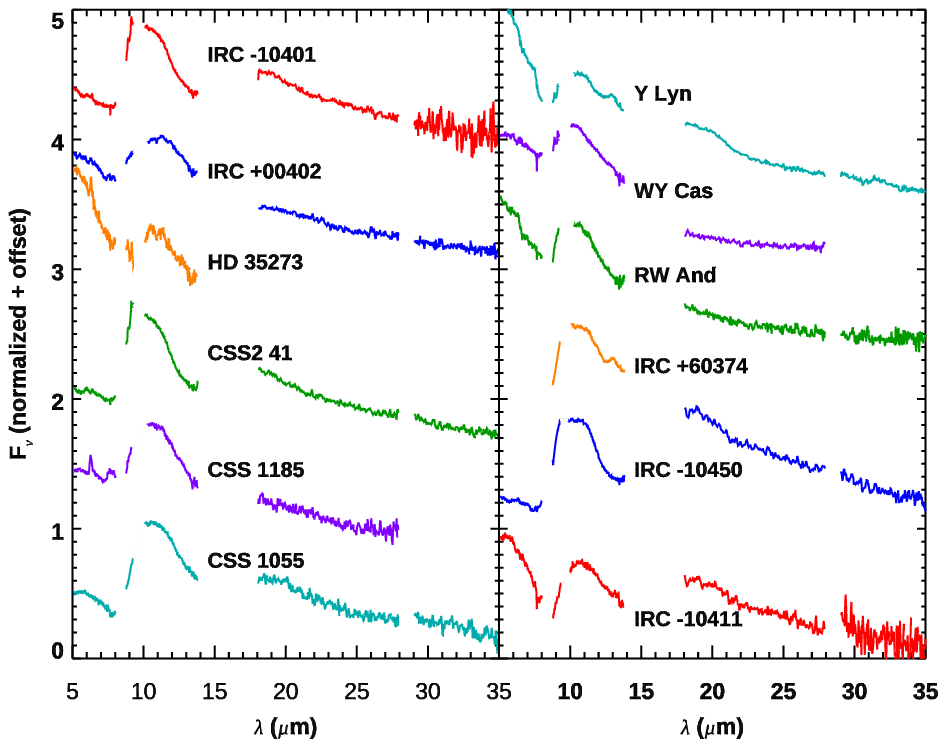}
\caption{SOFIA/FORCAST S star spectra. The data have been normalized at 
10.5--12 \mum\ and offset for clarity. The $\sim$9--10 \mum\ region is 
grayed-out due to residuals from atmospheric ozone, determined individually
for each star. A telluric residual is also sometimes present at 7.6--7.7 \mum.
}
\label{fig.all}
\end{figure*}

Figure \ref{fig.all} shows the FORCAST spectra for the observed
S stars. The spectra are dominated by a broad 
emission feature around 10 \mum, as expected for this kind of dust-producing 
star.  Several, but not all, show a weak 18 \mum\ feature similar to M giants.
Two stars also show the 13 \mum\ emission feature seen in some
O-rich AGB stars. At least two stars have an absorption feature from the 
\water\ band at 6.5 \mum, and at least three others have an emission feature 
from complex hydrocarbons near 6.3 \mum.
No SiS bands were detected in our sample. 
The detected gas and dust features were separately isolated from the continuum 
and characterized, as discussed below.

\subsection{Dust}\label{sec.dust}
\begin{figure*}[t] 
\includegraphics[width=6.5in]{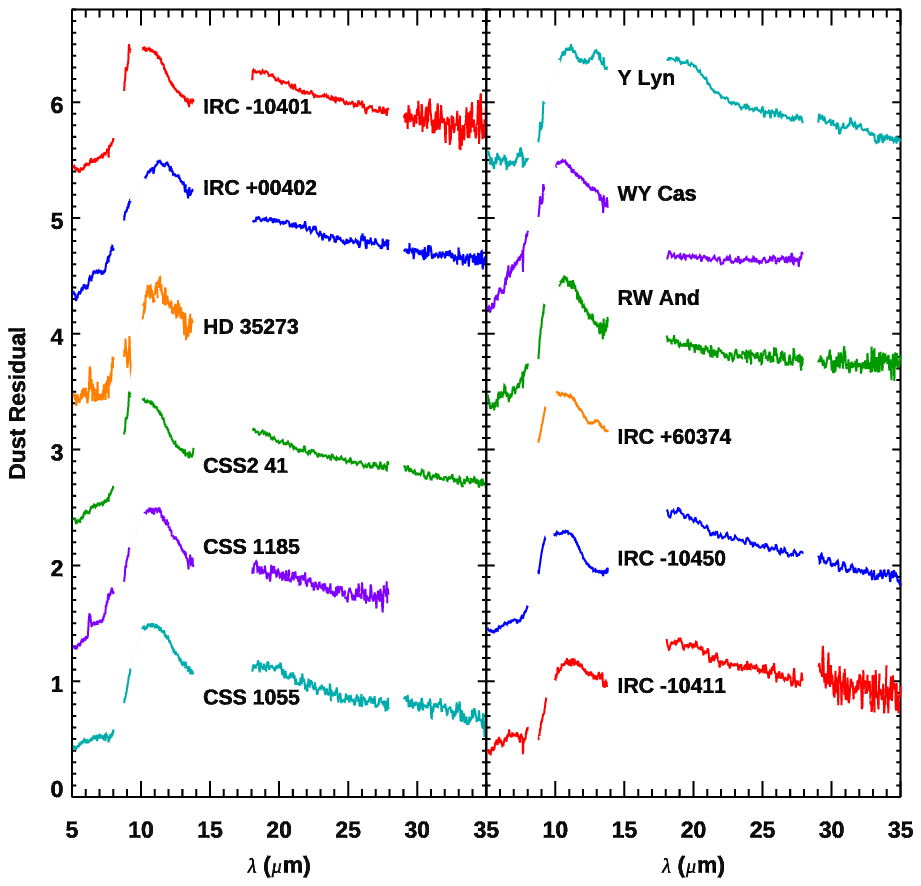}
\caption{Dust residuals. A stellar continuum was subtracted from each spectrum, 
the results normalized to the mean residual in G111, and offset for clarity.
}
\label{fig.dust}
\end{figure*}
\subsubsection{{Feature Extraction}}\label{sec.dust_extract}
To isolate the dust features, a stellar continuum was subtracted
from each spectrum. We used
the \spi\ IRS spectrum of CSS 454, a dust-free S star with relatively
weak molecular absorption \cite[][CSS 454 is the magenta asterisk and black square in Figure 
\ref{fig.colcol}]{smoldersea12a}\footnote{The bluest source in 
Figure \ref{fig.colcol}
from the previous samples is an extrinsic S star and thus not appropriate 
for a continuum template.}. The IRS spectrum was 
re-sampled to the FORCAST wavelength grid of each S star, scaled to the 
average flux density in the 5.5--7.8 \mum\ range, then subtracted from the 
target spectrum. 
Figure \ref{fig.dust} shows the resulting
residual spectra, which are dominated by the dust emission feature at $\sim$10 
\mum. 

S stars tend to be slightly O-rich due to the traditional identification
criteria\footnote{The traditional identification
from the ZrO absorption band in optical spectra \cite[e.g.,][]{merrill26zro, 
merrill29zro,wurm40} skews the known population slightly O-rich.}, 
so we 
compared the observed features to profiles from combinations of dust
commonly used oxygen-rich minerals, such as silicates, alumina, and gehlenite.
Because C/O is close to 1, we also included silicon carbide (SiC) 
and amorphous carbon.
Dust constants for the silicates were from \cite[][OHM]{ohm92},
the alumina constants from \citet[][Al$_2$O$_3$]{begemann+97}, the 
gehlenite \cite[Ca$_2$Al$_2$SiO$_7$, used by][for 
their S star sample]{smoldersea12a} were from \citet[][via the Jena Database of 
Optical Constants website\footnote{www.astro.uni-jena.de/Laboratory/Database/jpdoc/0-entry.html}]{mutschke+98}, 
 the silicon carbide from \cite{pegourie88}, and the amorphous carbon
from \cite{zubko+96}. A grain size of 0.1 
\mum\ was used to obtain absorption efficiencies (q) from the complex indices
of refraction provided by the references given above; the grain size for
the amorphous carbon was 0.3 \mum. We compared profiles 
that combined different amounts of OHM silicates, alumina, gehlenite, SiC,
and amorphous carbon to the observed dust features by eye. None of the 
comparison
profiles matched the data
particularly well, and the uniqueness of any ``match'' is questionable. 
A more complete library of minerals could potentially be used to 
formally fit to the observed dust features, but the need to consider 
both oxygen-
and carbon-rich minerals would make the uniqueness issue even worse.

We look at the shape of the G227 segment to determining whether 
or not there is an 18  \mum\ feature present in these spectra. In particular,
we look for inflections which could be caused by the red
shoulder of the feature, the blue shoulder being in the gap between G111 and
G227. Y Lyn shows clear inflections in this segment which likely 
indicate an 18 \mum\ feature. These inflections are weaker but probably 
present in four of the other stars and are noted in Table \ref{tab.dust}.
Other samples of S stars also usually have weak or absent 18 \mum\ features. 
When present, it often peaks closer to 19 \mum\ than the typical silicate 
feature at 18 \mum\ \cite[e.g.,][]{lml88}.

The dust residuals in Figure \ref{fig.dust} may contain
an underlying component whose lack of subtraction artificially enhances
the apparent strength of the emission in the 18 \mum\ region. This component 
is particularly evident in objects where the dust residuals are stronger 
above 17 \mum\
compared to those below 15 \mum, such as IRC $-$10450 and IRC $-$10411, but
is likely present in all our sources. \cite{smoldersea12a} included 
Planck functions with T=500--4000 K, as well as stellar spectra, to 
isolate their dust features, probably due to this emission excess, 
although they do not discuss the fitting 
process or resulting parameters. Possible carriers for the featureless 
excess emission could include amorphous carbon or iron grains 
\cite[e.g.,][]{mcd10}.
Another possibility is a poor calibration 
between FORCAST grism segments. However, Figure \ref{fig.all} shows that
the G227 and G329 segments are well aligned prior to subtraction of the 
stellar component, and G111 and G227 generally are, too. Comparison with
the IRAS flux densities at 12 and 25 \mum\ finds that sometimes the
values match the spectra, sometimes one does and not the other, and
sometimes neither does. This may be due to variability in the stars, but  
regardless does not help address the relative levels of the segments.

Two stars show the 13 \mum\ feature attributed to 
crystalline alumina \citep{sloan13um,takigawa+15}. As Figure \ref{fig.dust10} 
shows, Y Lyn has a strong 13 \mum\ 
feature, as well as the most distinct 18 \mum\ feature and the \water\
absorption band at 6.5 \mum\ (Sec. \ref{sec.water}). \cite{sloan13um} 
found that stars with a 13 \mum\ feature often showed a redder component
at $\sim$19.5--20 \mum\  to their 18 \mum\ feature (e.g., their Figure 1). 
While Y Lyn does not have a distinct emission bump at those wavelengths as
the oxygen-rich AGB stars often do, 
its ``18'' \mum\ inflection is redder than that of the other S stars 
with tentative 18 \mum\ features but no 13 \mum\ feature, such as CSS 1055.
IRC +60374, with
the other clear 13 \mum\ feature, only has data from the G111 grating, so it
is not known if it has \water\ absorption or an 18 \mum\ feature. 
WY Cas does have a hint of excess emission around 13 \mum, although 
it is much weaker than in Y Lyn or IRC +60374. IRC $-$10411 also has 
excess emission but at a longer wavelength than the typical 13 \mum\ feature.

\subsubsection{Dust Emission Classes}\label{sec.dustd}
\begin{figure} 
\includegraphics[width=3.35in]{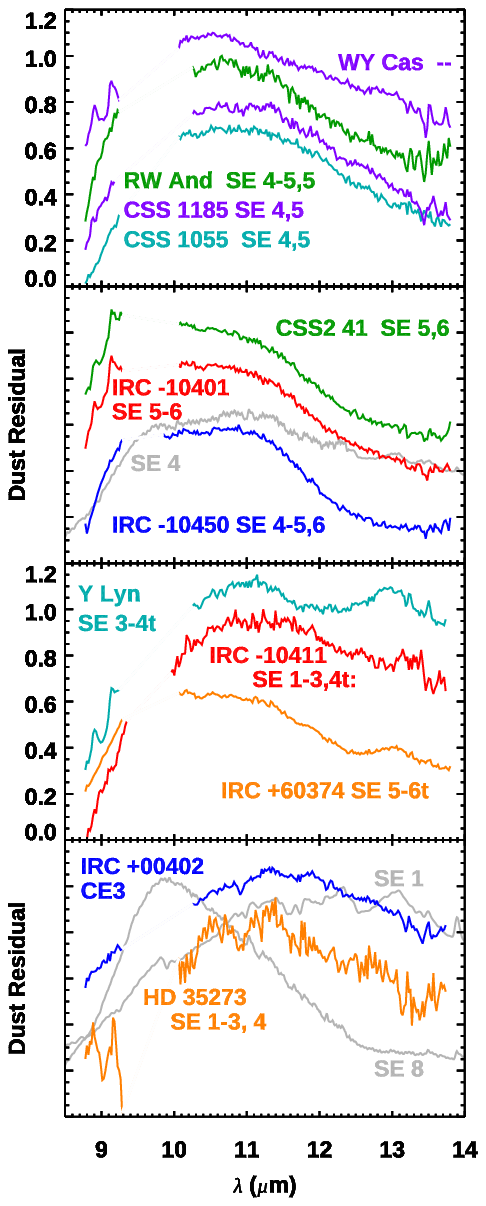}
\caption{Close-up of the dust features in the 10--11 \mum\ region. Roughly
similar spectra are grouped together. SE classes for the blue and 
red sides are separated by a comma if they differ. The sources with the 
13 \mum\ emission feature are in the third panel from the top. 
Spectra are normalized as described in the text and offset for clarity.
Comparison profiles for classes SE 1, 4, and 8 are in gray.}
\label{fig.dust10}
\end{figure}

The SE classes of \citet[][hereafter SP95 and SP98, respectively]{sp95,sp98} 
are commonly used to characterized the 10 \mum\
emission feature in O-rich AGB stars, and SP98 suggested that they could be
used on S stars. Generally, the spectra from classes with 
lower numbers, e.g., SE1--3, have more alumina and those with higher 
numbers have more amorphous silicates (SP98 and references therein).
Since the SP95 algorithm used to assign the class requires data within the 
ozone-contaminated region, though, we cannot formally assign an SE class 
to our spectra (see Eqns. 4--5 of SP98). Instead,
we compared the shape of the continuum-subtracted feature to the average
spectrum of each class shown by \cite{sloan13um} and assign a class by eye. 
These are given in Table \ref{tab.dust}, along with the approximate peak 
wavelength of the emission feature. Figure \ref{fig.dust10} shows a close-up
of the feature. Sources with (roughly) similar features are grouped together.

For several of the stars, the shape of the peak (somewhat) 
corresponds to one SE class, but the slope to the red, above $\sim$12 \mum, 
corresponds better to the next higher SE class. \cite{lml88} had found that 
the 10 \mum\ features of S stars were distinct from the O-rich AGB stars. 
\cite{honyea09} also found that the shapes of the 10 \mum\ features in
their S star sample differed from those of the O-rich AGB stars. Our
data support the suggestion by both groups that the dust composition
of the S stars likely differs from that of the O-rich AGB stars.

Two of the stars, IRC +00402 and WY Cas, do not correspond well to any of
the SE classes, albeit in different ways. This indicates that their mineralogy differs even more from the
standard alumina+classic silicates of O-rich AGB stars \cite[e.g.,][]{es01}, 
and that they may be relatively more carbon-rich than the others. 

IRC +00402 (Figure \ref{fig.dust10} lower-right panel) resembles some of the 
carbon
stars observed with SOFIA by \cite{ksk19} and would be classed as
a CE3 following \cite{slo16}. It is also one of our S stars with a hydrocarbon 
feature at 6.3 \mum, which normal carbon stars do not have.

The 10 \mum\ feature of WY Cas (Figure \ref{fig.dust10} top-left panel), 
though, does not resemble either the silicate
features of M giants or the SiC features of the carbon stars. It is also one
of the sources with no 18 \mum\ feature. Indeed it shows only weak emission in 
the G229 grating, and thus has less cool dust present compared to the other 
sources, regardless of the mineralogy that causes its odd 10 \mum\ feature.

\begin{deluxetable*}{lrrrrrrr}[ht]
\tablecaption{Feature Summary\label{tab.dust}}
\tablewidth{0pt}
\tablehead{
\colhead{Star} & \colhead{SE} & \colhead{$\lambda_{pk10}$} &
\colhead{18 \mum?} & \colhead{13 \mum?} & \colhead{\water} &
\colhead{$\lambda_{pk6}$} & \colhead{$\lambda_{c6}$}\\
\colhead{Name} & \colhead{Class} & \colhead{(\mum)} & & & & \colhead{(\mum)} 
& \colhead{(\mum)}
}
\startdata  
RW And      & 4-5, 5  & 10.7 & N       & N       & $\sim$3000 K & \nodata & \nodata  \\
HD 35273    & 1-3, 4  & 11.3 & \nodata & N       & \nodata      & 6.30    & 6.34     \\
Y Lyn       & 3-4t    & 11.0 & Y       & Y       & $\sim$3000 K & \nodata & \nodata  \\
IRC $-$10401 & 5-6     & $<$10.1   & N       & N       & \nodata      & 6.38    & 6.36     \\
IRC $-$10411 & 1-3, 4t:  & 11.2 & Y       & Y:       & \nodata      & \nodata & \nodata  \\
CSS 1055    & 4, 5    & 10.8 & Y       & N       & \nodata      & \nodata & \nodata  \\
IRC $-$10450 & 4-5, 6  & 10.6 & Y       & N       & weak         & \nodata & \nodata  \\
IRC +00402   & CE3     & 11.3 & ?       & N       & \nodata      & 6.30    & 6.32     \\
CSS2 41     & 5, 6    & $<$10.1   & N       & N       & \nodata      & \nodata & \nodata  \\
CSS 1185    & 4, 5    & 10.9 & N       & N       & \nodata      & 6.28    & 6.30     \\
IRC +60374   & 5-6t    & $<$10.2   & \nodata & Y       & \nodata      & \nodata & \nodata  \\
WY Cas      & \nodata & 10.5 & N       & N:       & \nodata      & \nodata & \nodata  
\enddata
\tablecomments{SE classes are given for the blue and red sides (if they 
differ), separated by a comma.  IRC +60374 
has no continuum subtraction, which could affect the
shape and peak wavelength for its 10 \mum\ feature.}
\end{deluxetable*}

\subsection{Molecular Absorption}\label{sec.water}
\begin{figure} 
\includegraphics[width=3.4in]{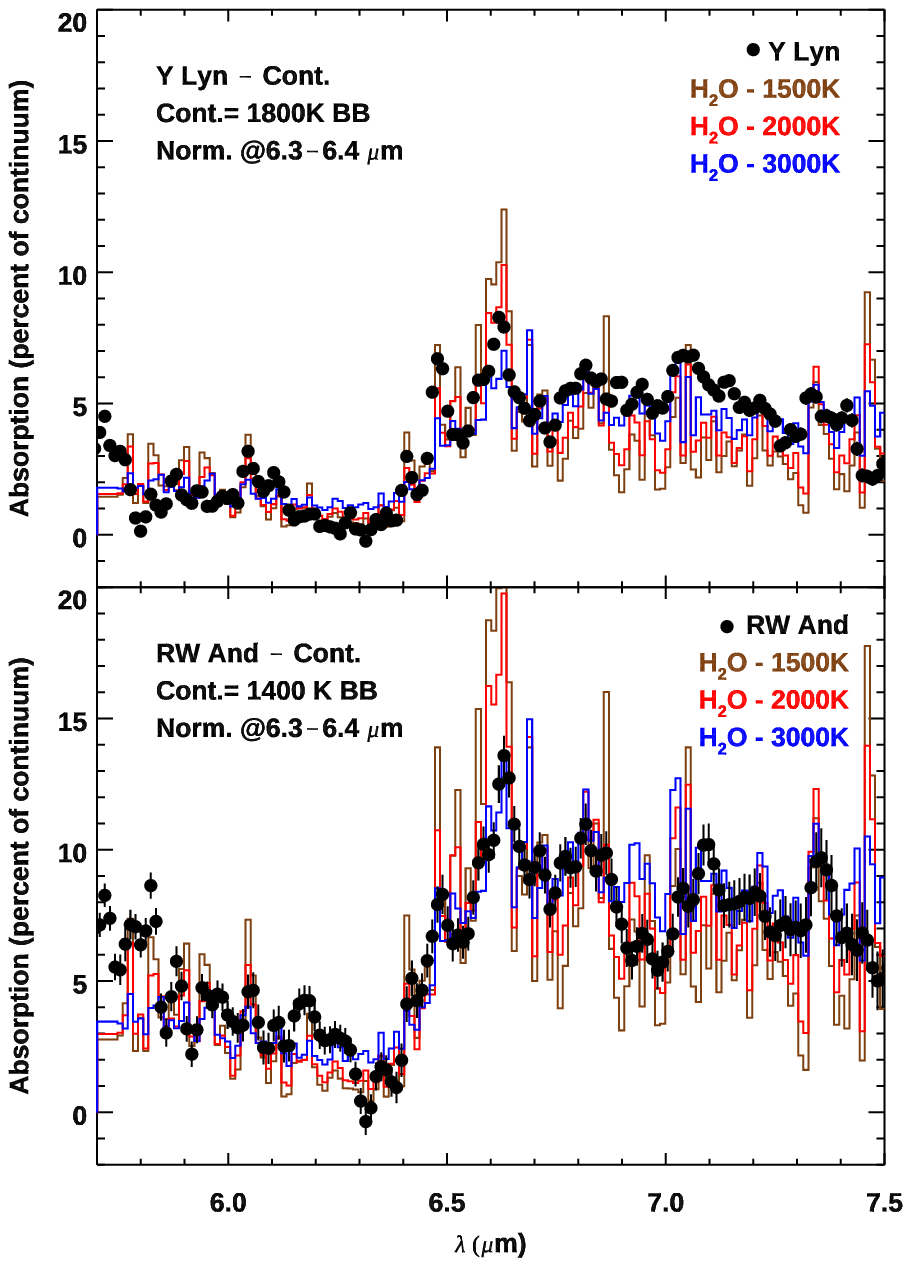}
\caption{\water\ models from HITEMP compared to (top) Y Lyn and (bottom) RW And.
The subtracted continuum is a blackbody function at the given 
temperature.}
\label{fig.water}
\end{figure}

Two of the S stars, Y Lyn and RW And, have a clear \water\ absorption 
bands in the 6.5--7.5 
\mum\ range. A third star, IRC $-$10450, has a tentative 
detection but is too weak to characterize.
To isolate the feature in Y Lyn and RW And, we subtract a Planck function
from the data. The best temperature was
T$_{\rm bb}$=1800 K and 1400 K for Y Lyn and RW And, respectively, scaled to 
the measurements at 6.3--6.4 \mum. The result was then converted into 
absorption as a percentage of the continuum for comparison to the \water\ 
models. 
The \water\ absorption models are based on the HITEMP extension of the HITRAN 
database \citep{hitemp95, hitemp2010}, using KSPECTRUM \citep{kop+13, rl18}. 
They were generated at 1500, 2000, and 3000 K, and Figure \ref{fig.water} 
shows the results. For both stars, 
the 3000 K model matches the data somewhat better than the cooler models. 
Choosing a different temperature or normalization range will change the 
details of the agreement, but in most cases the higher temperature remains 
the better
match compared to the lower-temperature water across most of the wavelength
range considered. 

This temperature contrasts with the \water\ observed in a set of 
dust-free M giants, which were better fit by the lower-temperature 
models, 1500 K and 2000 K \citep{sloanea15}. A higher temperature could indicate that
the absorbing gas is closer to the stellar surface in the S stars. The 
different abundances of the available oxygen could affect where in the 
atmosphere the \water\ can condense. Given the noisiness of the spectra 
and the uncertainty of the fit, this statement should be taken more as guidance
than a strong conclusion.

The edge of the SiO absorption band at 7.5 \mum\ 
was detected in Y Lyn but not in RW And (Figure \ref{fig.all}). Y Lyn does have 
the bluest
K$_s-$[12] color in our sample, although not the bluest [12]$-$[25].
All but one of the naked (i.e., dust-free) stars in the 
\cite{smoldersea12a} sample show SiO, and many also have \water\ absorption
(their Figure 4). One star, CSS 783, may have weak \water\ absorption with no 
SiO feature.
Of the dusty S stars in their sample (their Figures 6 and 7),
they typically have either both absorption features or neither, and a few only
have SiO. Here, too, a single source, CD $-$392449, may have weak \water\ 
absorption without the SiO feature.

Thus, RW And seems unusual in having clear \water\ absorption but not SiO 
absorption. Two additional stars in the Smolders sample may also show
this unusual combination but higher signal-to-noise data are needed to 
verify it. It is possible that a lack of available oxygen means that the
formation of dust has consumed most of the SiO gas, leaving an insufficient
amount for the absorption feature to appear in these stars.

\subsection{Complex Hydrocarbon Features}\label{sec.pahs}
Infrared emission features attributed to complex hydrocarbons have 
frequently been observed in carbon-rich evolved objects such as 
post-AGB stars and planetary nebulae, typically at wavelengths of 
$\sim$3.3, 6.2, 7.7-8.6, and 11.3 \mum.  The features are usually 
associated with PAHs and related aliphatics \cite[e.g.,][]{dw81,tielens08} 
although the detailed composition of their carrier(s) remains the 
topic of some debate \cite[e.g.,][]{kwok22}. They 
 are generally not seen in carbon stars, as opposed to carbon-rich objects
that have evolved beyond the AGB, due to the lack of UV photons to excite
the carriers \cite[e.g.,][]{tielens08}. Nonetheless, the features
have been detected in cool objects such as red supergiants 
\cite[e.g.,][]{sylvesterea94, sylvesterea98, verhoelstea09, jonesea17}.

\subsubsection{Feature Extraction}
\begin{figure} 
\includegraphics[width=3.4in]{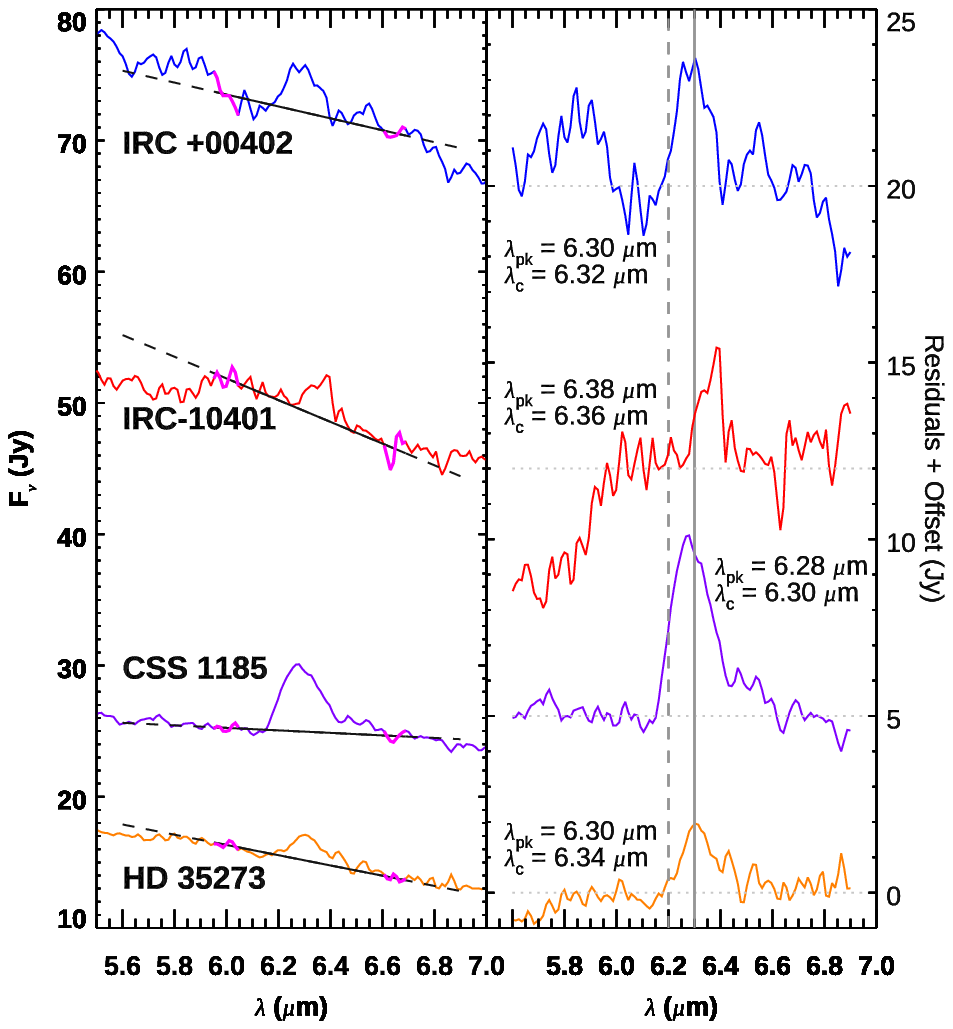}
\caption{Candidate hydrocarbon features. (left) Continuum fits to the four S stars with
the 6.3 \mum\ hydrocarbon feature. (right) Residuals after continuum subtraction, 
offset for clarity. The vertical lines are at 6.2 and 6.3 \mum\ to guide the eye. 
}
\label{fig.uir6}
\end{figure}

Three S stars show the hydrocarbon feature at 6.3 \mum,  CSS 1185, HD 35273, and
IRC +00402. In addition, IRC $-$10401 has a candidate feature.
To remove the continuum and isolate the feature, we fit a line to the mean 
flux levels between 5.95 and 6.05 \mum\ on the blue side
and between 6.60 and 6.70 \mum\ on the red side and subtract that from the 
spectrum, as shown in the left panel of Figure \ref{fig.uir6}.  The right panel 
in the figure shows the residual features for the four stars. 

The wavelength at which the feature appears depends slightly on how it is 
defined. Table \ref{tab.dust} gives both $\lambda_{pk}$, the peak
wavelength and $\lambda_{c}$ the wavelength centroid, for each source. The
peak wavelength is the wavelength of the peak flux in the continuum-subtracted
feature. The wavelength centroid is the wavelength at which half the flux in
the continuum-subtracted feature is on either side. These can shift slightly,
$\sim$0.01--0.02 \mum, depending on the wavelengths chosen, which
indicates the uncertainty in the reported wavelengths.

For three of the four stars, the feature appears at 
$\langle\lambda\rangle$=6.31$\pm$0.02 \mum.  This corresponds to ``Class C'' 
PAHs, 
the reddest in the classification scheme of \cite{peetersea02} based on
spectra from the Short-Wavelength Spectrometer \citep[SWS;][]{sws96} 
on the Infrared Space Observatory \citep[ISO][]{iso96}.
The candidate feature in IRC $-$10401 is even redder, at $\sim$6.37 \mum. 

\begin{figure} 
\includegraphics[width=3.4in]{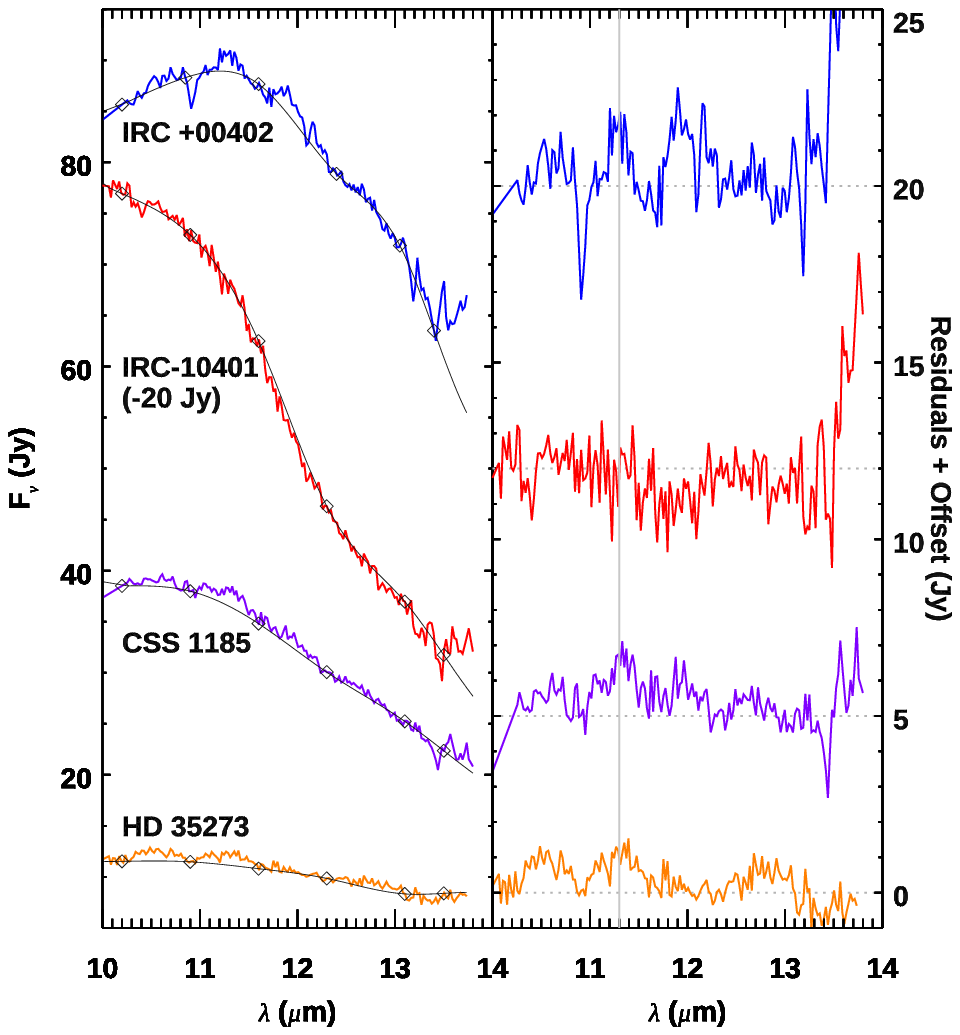}
\caption{The 11 \mum\ region. (left) Continuum spline fits for the four stars 
with 6.3 \mum\ features. (right) Residuals after continuum subtraction. 
The vertical gray line is at 11.3 \mum.}
\label{fig.uir11}
\end{figure}

The 11.3 \mum\ feature is also commonly observed in spectra with complex
hydrocarbon features,
including some S stars \citep{smoldersea10}. As Figure \ref{fig.uir11} shows, 
though, none of the four spectra with 6.3 \mum\ features has a clear 
emission feature at 11.3 \mum. In this spectral region, we used a spline 
to try and bring out the feature. There is a hint
of emission in the three stars with positively identified 6.3 \mum\ features, 
but that can
only be considered tentative; there is no sign of a similar feature for 
IRC $-$10401. 

The 7.7--8.6 \mum\ complex lies almost entirely in the gap between the G063
and G111 gratings. The spectra for our PAH sources, especially CSS 1185, do
show some structure around 7.3--8.0 \mum\ (see Figure \ref{fig.all} 
or \ref{fig.dust}). However, the wavelength coverage is insufficient 
 to characterize it, or really to distinguish 
between hydrocarbon emission and the 
silicate dust feature that is certainly present and beginning to rise. 

\subsubsection{Hydrocarbon Excitation}\label{sec.pahsd}

Class C features, at 6.3 \mum, rather than 6.22--6.28 \mum, were originally 
found in only two carbon-rich post-AGB objects \citep{peetersea02}. 
Although they have been found in other object types \cite[][and references 
therein]{sloanea14},
they remain rare compared to the Class A and B sources. 

Class C objects generally show a single, broad feature 
$\sim$8.0--8.3 \mum\ rather than two distinct features at 7.7 and 8.6 \mum\
\cite[e.g.,][]{peetersea02, jensen+22}. \cite{sjd07} note that the 8.6 
\mum\ feature does often
appear as a bump on the red shoulder of the broader feature in Class C spectra.
The 11.3 \mum\ feature is also redder than in Classes A and B.

Three of the four hydrocarbon sources among the S stars of \cite{smoldersea10} 
are also Class C, with a red 6.3 \mum\ feature and broad 8 \mum\ 
features. Their fourth source has an 11.3 \mum\ feature, but it is Class B in
all its PAH features. That is, at least six of the eight S stars with 
hydrocarbon features are Class C. The candidate 6.3 \mum\ feature in the
eighth source, if real, is even redder than Class C. 

Class C features are thought to indicate a lack of photo-processing
of the molecules. 
\cite{sjd07} suggested that the wavelength shift from 6.2 
to 6.3 \mum\ could be due to the influence of aliphatic bonds relative to 
aromatics. Laboratory measurements of soots by \cite{pino+08} support this 
scenario. In astrophysical environments, the aliphatics could survive 
due to the absence of UV radiation, which would certainly be the
case for these cool S stars, T$_{\star}\sim$2000--3000 K. 

The ratio of the 6.2/11.3 \mum\ feature is often used to 
estimate the charge state of the PAHs, as the 6.2 \mum\ feature is 
attributed to cations and the 11.3 feature to neutrals 
\cite[e.g.,][and references therein]{peeters+17}. A weak or absent 11.3 
\mum\ feature  could indicate high ionization. However, that would 
conflict with the lack of UV processing indicated by the wavelength 
shift of the 6.2 \mum\ feature to 6.3 \mum. Since a lack of UV radiation is
more consistent with the stellar temperatures of S stars, a different
explanation for the lack of the 11.3 \mum\ feature is needed here.

The 11.3 \mum\ feature is attributed to a solo C-H out-of-plane bending 
mode in large, neutral PAHs. An alternate explanation for the absent
feature in the S stars is an overabundance of aliphatics rather than 
aromatics. In this scenario, the hydrogen atoms on the edges of the 
hydrocarbons are replaced by chains such as methyl or methylene groups. 
One might expect such a replacement in an aliphatic-rich hydrocarbon 
mixture, and this could have the effect of suppressing the solo C-H 
out-of-plane bending mode. 

A test of this might be found in the 3--4 \mum\ spectral region, which
covers both the aromatic feature at 3.3 \mum\ and the aliphatic
feature at 3.4 \mum. None of the S stars with the 6.3 \mum\ feature 
in our sample, though, or that of \cite{smoldersea10}, have 3 \mum\ spectra.
The SWS spectra of the two original Class C post-AGB objects, AFGL 
2688 and IRAS 13416$-$6243, do extend to 2.4 \mum. However, the signal-to-noise
ratio in these data is insufficient for a definitive test of this scenario.
Spectral observations at 3 \mum\ of the S stars 
themselves are needed to determine the relative abundances of aromatics and
aliphatics in these objects.

\section{Summary} \label{sec.summ}

We observed 12 of the dustiest known S stars in 
the Milky Way using the FORCAST grisms on SOFIA. All have strong 
 dust emission features in the vicinity of the 10 \mum\ and 11 \mum\
silicate and SiC features.
The shape of the feature in most of the stars, though, is not well-matched 
to the shapes for a single SE class in the scheme often used to categorize dusty
oxygen-rich AGB stars. This supports the contention that the dust emission 
from S stars differs from that in M giants \cite[e.g.,][]{lml88,honyea09}. 
The feature for one star (IRC +00402) is likely due to 
silicon carbide, indicating its C/O is slightly above 1, and the feature for
another star (WY Cas) does not resemble either typical M giant features nor 
carbon star features.

Two stars show water absorption
band $\sim$6.5--7.5 \mum, a third has a tentative detection, but only
one of these also has SiO molecular absorption. This combination of water 
absorption without SiO absorption has only been tentatively seen in two 
other S stars, and is never seen in M giants.

Three stars 
show the 6.3 \mum\ emission band from complex hydrocarbons, and a fourth may 
have emission at 6.37 \mum, redder than even the class C
feature the other three show. None of them shows a strong 11.3 \mum\ 
hydrocarbon emission feature, 
although the three class C sources have tentative detections. The 
combination of a red 6.3 \mum\ feature and weak 11.3 \mum\ feature may indicate
that aliphatic hydrocarbons are relatively more abundant in these sources 
compared to the PAHs commonly seen in nebulae. 
Follow-up observations in the 3.3 \mum\ spectral region would test this
possible explanation.

\begin{acknowledgments}

We thank the SOFIA flight crew
and staff scientists for making these observations, Els Peeters for
helpful discussions on the hydrocarbon emission, and Tom Kuchar for assisting 
with the photometry. We thank the anonymous referee whose careful reading 
and thoughtful suggestions have helped improve the paper. 
Based on observations made with the NASA/DLR Stratospheric
Observatory for Infrared Astronomy (SOFIA). SOFIA was jointly operated by the
Universities Space Research Association, Inc. (USRA), under NASA contract
NNA17BF53C, and the Deutsches SOFIA Institut (DSI) under DLR contract
50 OK 0901 to the University of Stuttgart. Financial support for this work
was provided by NASA through award SOF09-0046 issued by USRA.
We made use of the
NASA Astrophysics Data System, IRSA's Gator service, and CDS's Simbad and
Vizier services.
\end{acknowledgments}

\vspace{3mm}

\facilities{SOFIA (FORCAST), \spi~(IRS), \iso~(SWS)}

\appendix
\section{S Stars Not Observed}
Table \ref{tab.notobs} lists the four S stars in our sample that were not 
observed by the end of the SOFIA mission.

\begin{deluxetable*}{lrrrrrr}[ht]
\tablecaption{Non-Observed S Stars\label{tab.notobs}}
\tablewidth{0pt}
\tablehead{
\colhead{Star} & 
\colhead{[12]$-$[25] }  & \colhead{K$_s$$-$[12]}& \colhead{Spectral} \\
\colhead{Name} 
& \colhead{(mag)} & \colhead{(mag)}& \colhead{Type}
}
\startdata
TT CMa    &  0.95 & 1.94 & S \\
CGCS 4284 &  1.03 & 4.59 & S \\
AFGL 2425 &  1.37 & 3.57 & M10 III \\
V1959 Cyg &  1.30 & 2.83 & S\\
\enddata
\end{deluxetable*}

\bibliography{sstars-refs_rev1}{}

\begin{thebibliography}{}
\expandafter\ifx\csname natexlab\endcsname\relax\def\natexlab#1{#1}\fi
\providecommand{\url}[1]{\href{#1}{#1}}
\providecommand{\dodoi}[1]{doi:~\href{http://doi.org/#1}{\nolinkurl{#1}}}
\providecommand{\doeprint}[1]{\href{http://ascl.net/#1}{\nolinkurl{http://ascl.net/#1}}}
\providecommand{\doarXiv}[1]{\href{https://arxiv.org/abs/#1}{\nolinkurl{https://arxiv.org/abs/#1}}}

\bibitem[{{Aoki} {et~al.}(1998){Aoki}, {Tsuji}, \& {Ohnaka}}]{aokiea98}
{Aoki}, W., {Tsuji}, T., \& {Ohnaka}, K. 1998, \aap, 340, 222

\bibitem[{{Begemann} {et~al.}(1997){Begemann}, {Dorschner}, {Henning},
  {Mutschke}, {G{\"u}rtler}, {K{\"o}mpe}, \& {Nass}}]{begemann+97}
{Begemann}, B., {Dorschner}, J., {Henning}, T., {et~al.} 1997, \apj, 476, 199,
  \dodoi{10.1086/303597}

\bibitem[{{Burbidge} {et~al.}(1957){Burbidge}, {Burbidge}, {Fowler}, \&
  {Hoyle}}]{burbidge+57}
{Burbidge}, E.~M., {Burbidge}, G.~R., {Fowler}, W.~A., \& {Hoyle}, F. 1957,
  Reviews of Modern Physics, 29, 547, \dodoi{10.1103/RevModPhys.29.547}

\bibitem[{{Cami} {et~al.}(2009){Cami}, {Sloan}, {Markwick-Kemper}, {Zijlstra},
  {Bauschlicher}, {Matsuura}, {Decin}, \& {Hony}}]{camiea09}
{Cami}, J., {Sloan}, G.~C., {Markwick-Kemper}, A.~J., {et~al.} 2009, \apjl,
  690, L122, \dodoi{10.1088/0004-637X/690/2/L122}

\bibitem[{{Cheeseman} {et~al.}(1989){Cheeseman}, {Stutz}, {Self}, {Taylor},
  {Goebel}, {Volk}, \& {Walker}}]{cheesemanea89}
{Cheeseman}, P., {Stutz}, J., {Self}, M., {et~al.} 1989, NASA Reference
  Publication, 1217, 1

\bibitem[{{Chen} {et~al.}(2019){Chen}, {Liu}, \& {Shan}}]{chenea19}
{Chen}, P.~S., {Liu}, J.~Y., \& {Shan}, H.~G. 2019, \aj, 158, 22,
  \dodoi{10.3847/1538-3881/ab2334}

\bibitem[{{de Graauw} {et~al.}(1996){de Graauw}, {Haser}, {Beintema},
  {Roelfsema}, {van Agthoven}, {Barl}, {Bauer}, {Bekenkamp}, {Boonstra},
  {Boxhoorn}, {Cote}, {de Groene}, {van Dijkhuizen}, {Drapatz}, {Evers},
  {Feuchtgruber}, {Frericks}, {Genzel}, {Haerendel}, {Heras}, {van der Hucht},
  {van der Hulst}, {Huygen}, {Jacobs}, {Jakob}, {Kamperman}, {Katterloher},
  {Kester}, {Kunze}, {Kussendrager}, {Lahuis}, {Lamers}, {Leech}, {van der
  Lei}, {van der Linden}, {Luinge}, {Lutz}, {Melzner}, {Morris}, {van Nguyen},
  {Ploeger}, {Price}, {Salama}, {Schaeidt}, {Sijm}, {Smoorenburg}, {Spakman},
  {Spoon}, {Steinmayer}, {Stoecker}, {Valentijn}, {Vandenbussche}, {Visser},
  {Waelkens}, {Waters}, {Wensink}, {Wesselius}, {Wiezorrek}, {Wieprecht},
  {Wijnbergen}, {Wildeman}, \& {Young}}]{sws96}
{de Graauw}, T., {Haser}, L.~N., {Beintema}, D.~A., {et~al.} 1996, \aap, 315,
  L49

\bibitem[{{Duley} \& {Williams}(1981)}]{dw81}
{Duley}, W.~W., \& {Williams}, D.~A. 1981, \mnras, 196, 269,
  \dodoi{10.1093/mnras/196.2.269}

\bibitem[{{Egan} \& {Sloan}(2001)}]{es01}
{Egan}, M.~P., \& {Sloan}, G.~C. 2001, \apj, 558, 165, \dodoi{10.1086/322443}

\bibitem[{{Habing}(1996)}]{hab96}
{Habing}, H.~J. 1996, \aapr, 7, 97, \dodoi{10.1007/PL00013287}

\bibitem[{{Hackwell}(1972)}]{hackwell72}
{Hackwell}, J.~A. 1972, \aap, 21, 239

\bibitem[{{Herter} {et~al.}(2012){Herter}, {Adams}, {De Buizer}, {Gull},
  {Schoenwald}, {Henderson}, {Keller}, {Nikola}, {Stacey}, \&
  {Vacca}}]{forcast}
{Herter}, T.~L., {Adams}, J.~D., {De Buizer}, J.~M., {et~al.} 2012, \apjl, 749,
  L18, \dodoi{10.1088/2041-8205/749/2/L18}

\bibitem[{{H{\"o}fner} \& {Olofsson}(2018)}]{ho18}
{H{\"o}fner}, S., \& {Olofsson}, H. 2018, \aapr, 26, 1,
  \dodoi{10.1007/s00159-017-0106-5}

\bibitem[{{Hony} {et~al.}(2009){Hony}, {Heras}, {Molster}, \&
  {Smolders}}]{honyea09}
{Hony}, S., {Heras}, A.~M., {Molster}, F.~J., \& {Smolders}, K. 2009, \aap,
  501, 609, \dodoi{10.1051/0004-6361/200912017}

\bibitem[{{Iben} \& {Renzini}(1983)}]{ir83}
{Iben}, Jr., I., \& {Renzini}, A. 1983, \araa, 21, 271,
  \dodoi{10.1146/annurev.aa.21.090183.001415}

\bibitem[{{Iben}(1974)}]{iben74}
{Iben}, I., J. 1974, \araa, 12, 215,
  \dodoi{10.1146/annurev.aa.12.090174.001243}

\bibitem[{{Jensen} {et~al.}(2022){Jensen}, {Shannon}, {Peeters}, {Sloan}, \&
  {Stock}}]{jensen+22}
{Jensen}, P.~A., {Shannon}, M.~J., {Peeters}, E., {Sloan}, G.~C., \& {Stock},
  D.~J. 2022, \aap, 665, A153, \dodoi{10.1051/0004-6361/202141511}

\bibitem[{{Jones} {et~al.}(2017){Jones}, {Woods}, {Kemper}, {Kraemer}, {Sloan},
  {Srinivasan}, {Oliveira}, {van Loon}, {Boyer}, {Sargent}, {McDonald},
  {Meixner}, {Zijlstra}, {Ruffle}, {Lagadec}, {Pauly}, {Sewi{\l}o}, {Clayton},
  \& {Volk}}]{jonesea17}
{Jones}, O.~C., {Woods}, P.~M., {Kemper}, F., {et~al.} 2017, \mnras, 470, 3250,
  \dodoi{10.1093/mnras/stx1101}

\bibitem[{{Karakas} \& {Lattanzio}(2014)}]{kl14}
{Karakas}, A.~I., \& {Lattanzio}, J.~C. 2014, \pasa, 31, e030,
  \dodoi{10.1017/pasa.2014.21}

\bibitem[{{Kessler} {et~al.}(1996){Kessler}, {Steinz}, {Anderegg}, {Clavel},
  {Drechsel}, {Estaria}, {Faelker}, {Riedinger}, {Robson}, {Taylor}, \&
  {Xim{\'e}nez de Ferr{\'a}n}}]{iso96}
{Kessler}, M.~F., {Steinz}, J.~A., {Anderegg}, M.~E., {et~al.} 1996, \aap, 315,
  L27

\bibitem[{{Kobayashi} {et~al.}(2020){Kobayashi}, {Karakas}, \&
  {Lugaro}}]{kobayashi+20}
{Kobayashi}, C., {Karakas}, A.~I., \& {Lugaro}, M. 2020, \apj, 900, 179,
  \dodoi{10.3847/1538-4357/abae65}

\bibitem[{{Kopparapu} {et~al.}(2013){Kopparapu}, {Ramirez}, {Kasting}, {Eymet},
  {Robinson}, {Mahadevan}, {Terrien}, {Domagal-Goldman}, {Meadows}, \&
  {Deshpande}}]{kop+13}
{Kopparapu}, R.~K., {Ramirez}, R., {Kasting}, J.~F., {et~al.} 2013, \apj, 765,
  131, \dodoi{10.1088/0004-637X/765/2/131}

\bibitem[{{Kraemer} {et~al.}(2019){Kraemer}, {Sloan}, {Keller}, {McDonald},
  {Zijlstra}, \& {Groenewegen}}]{ksk19}
{Kraemer}, K.~E., {Sloan}, G.~C., {Keller}, L.~D., {et~al.} 2019, \apj, 887,
  82, \dodoi{10.3847/1538-4357/ab4f6b}

\bibitem[{{Kraemer} {et~al.}(2002){Kraemer}, {Sloan}, {Price}, \&
  {Walker}}]{kspw02}
{Kraemer}, K.~E., {Sloan}, G.~C., {Price}, S.~D., \& {Walker}, H.~J. 2002,
  \apjs, 140, 389, \dodoi{10.1086/339708}

\bibitem[{{Kwok}(2022)}]{kwok22}
{Kwok}, S. 2022, \apss, 367, 16, \dodoi{10.1007/s10509-022-04045-6}

\bibitem[{{Little-Marenin} \& {Little}(1988)}]{lml88}
{Little-Marenin}, I.~R., \& {Little}, S.~J. 1988, \apj, 333, 305,
  \dodoi{10.1086/166747}

\bibitem[{{Matsuura} {et~al.}(2005){Matsuura}, {Zijlstra}, {van Loon},
  {Yamamura}, {Markwick}, {Whitelock}, {Woods}, {Marshall}, {Feast}, \&
  {Waters}}]{matsuuraea05}
{Matsuura}, M., {Zijlstra}, A.~A., {van Loon}, J.~T., {et~al.} 2005, \aap, 434,
  691, \dodoi{10.1051/0004-6361:20042305}

\bibitem[{{McDonald} {et~al.}(2010){McDonald}, {Sloan}, {Zijlstra},
  {Matsunaga}, {Matsuura}, {Kraemer}, {Bernard-Salas}, \& {Markwick}}]{mcd10}
{McDonald}, I., {Sloan}, G.~C., {Zijlstra}, A.~A., {et~al.} 2010, \apjl, 717,
  L92, \dodoi{10.1088/2041-8205/717/2/L92}

\bibitem[{{Merrill} \& {Stein}(1976)}]{merrillstein76}
{Merrill}, K.~M., \& {Stein}, W.~A. 1976, \pasp, 88, 285,
  \dodoi{10.1086/129945}

\bibitem[{{Merrill}(1926)}]{merrill26zro}
{Merrill}, P.~W. 1926, \apj, 63, 13, \dodoi{10.1086/142946}

\bibitem[{{Merrill}(1929)}]{merrill29zro}
---. 1929, Popular Astronomy, 37, 444

\bibitem[{{Mutschke} {et~al.}(1998){Mutschke}, {Begemann}, {Dorschner},
  {Guertler}, {Gustafson}, {Henning}, \& {Stognienko}}]{mutschke+98}
{Mutschke}, H., {Begemann}, B., {Dorschner}, J., {et~al.} 1998, \aap, 333, 188

\bibitem[{{Neugebauer} {et~al.}(1984){Neugebauer}, {Habing}, {van Duinen},
  {Aumann}, {Baud}, {Beichman}, {Beintema}, {Boggess}, {Clegg}, {de Jong},
  {Emerson}, {Gautier}, {Gillett}, {Harris}, {Hauser}, {Houck}, {Jennings},
  {Low}, {Marsden}, {Miley}, {Olnon}, {Pottasch}, {Raimond}, {Rowan-Robinson},
  {Soifer}, {Walker}, {Wesselius}, \& {Young}}]{iras84}
{Neugebauer}, G., {Habing}, H.~J., {van Duinen}, R., {et~al.} 1984, \apjl, 278,
  L1, \dodoi{10.1086/184209}

\bibitem[{{Ossenkopf} {et~al.}(1992){Ossenkopf}, {Henning}, \&
  {Mathis}}]{ohm92}
{Ossenkopf}, V., {Henning}, T., \& {Mathis}, J.~S. 1992, \aap, 261, 567

\bibitem[{{Peeters} {et~al.}(2017){Peeters}, {Bauschlicher}, {Allamandola},
  {Tielens}, {Ricca}, \& {Wolfire}}]{peeters+17}
{Peeters}, E., {Bauschlicher}, Charles~W., J., {Allamandola}, L.~J., {et~al.}
  2017, \apj, 836, 198, \dodoi{10.3847/1538-4357/836/2/198}

\bibitem[{{Peeters} {et~al.}(2002){Peeters}, {Hony}, {Van Kerckhoven},
  {Tielens}, {Allamandola}, {Hudgins}, \& {Bauschlicher}}]{peetersea02}
{Peeters}, E., {Hony}, S., {Van Kerckhoven}, C., {et~al.} 2002, \aap, 390,
  1089, \dodoi{10.1051/0004-6361:20020773}

\bibitem[{{Pegourie}(1988)}]{pegourie88}
{Pegourie}, B. 1988, \aap, 194, 335

\bibitem[{{Pino} {et~al.}(2008){Pino}, {Dartois}, {Cao}, {Carpentier},
  {Chamaill{\'e}}, {Vasquez}, {Jones}, {D'Hendecourt}, \&
  {Br{\'e}chignac}}]{pino+08}
{Pino}, T., {Dartois}, E., {Cao}, A.~T., {et~al.} 2008, \aap, 490, 665,
  \dodoi{10.1051/0004-6361:200809927}

\bibitem[{{Ramirez} \& {Levi}(2018)}]{rl18}
{Ramirez}, R.~M., \& {Levi}, A. 2018, \mnras, 477, 4627,
  \dodoi{10.1093/mnras/sty761}

\bibitem[{{Rothman} {et~al.}(1995){Rothman}, {Wattson}, {Gamache}, {Schroeder},
  \& {McCann}}]{hitemp95}
{Rothman}, L.~S., {Wattson}, R.~B., {Gamache}, R., {Schroeder}, J.~W., \&
  {McCann}, A. 1995, in Society of Photo-Optical Instrumentation Engineers
  (SPIE) Conference Series, Vol. 2471, Atmospheric Propagation and Remote
  Sensing IV, ed. J.~C. {Dainty}, 105--111, \dodoi{10.1117/12.211919}

\bibitem[{{Rothman} {et~al.}(2010){Rothman}, {Gordon}, {Barber}, {Dothe},
  {Gamache}, {Goldman}, {Perevalov}, {Tashkun}, \& {Tennyson}}]{hitemp2010}
{Rothman}, L.~S., {Gordon}, I.~E., {Barber}, R.~J., {et~al.} 2010, \jqsrt, 111,
  2139, \dodoi{10.1016/j.jqsrt.2010.05.001}

\bibitem[{{Ruffle} {et~al.}(2015){Ruffle}, {Kemper}, {Jones}, {Sloan},
  {Kraemer}, {Woods}, {Boyer}, {Srinivasan}, {Antoniou}, {Lagadec}, {Matsuura},
  {McDonald}, {Oliveira}, {Sargent}, {Sewi{\l}o}, {Szczerba}, {van Loon},
  {Volk}, \& {Zijlstra}}]{ruffleea15}
{Ruffle}, P.~M.~E., {Kemper}, F., {Jones}, O.~C., {et~al.} 2015, \mnras, 451,
  3504, \dodoi{10.1093/mnras/stv1106}

\bibitem[{{Sloan} {et~al.}(2015){Sloan}, {Goes}, {Ramirez}, {Kraemer}, \&
  {Engelke}}]{sloanea15}
{Sloan}, G.~C., {Goes}, C., {Ramirez}, R.~M., {Kraemer}, K.~E., \& {Engelke},
  C.~W. 2015, \apj, 811, 45, \dodoi{10.1088/0004-637X/811/1/45}

\bibitem[{{Sloan} {et~al.}(2003){Sloan}, {Kraemer}, {Goebel}, \&
  {Price}}]{sloan13um}
{Sloan}, G.~C., {Kraemer}, K.~E., {Goebel}, J.~H., \& {Price}, S.~D. 2003,
  \apj, 594, 483, \dodoi{10.1086/376857}

\bibitem[{{Sloan} \& {Price}(1995)}]{sp95}
{Sloan}, G.~C., \& {Price}, S.~D. 1995, \apj, 451, 758, \dodoi{10.1086/176262}

\bibitem[{{Sloan} \& {Price}(1998)}]{sp98}
---. 1998, \apjs, 119, 141, \dodoi{10.1086/313156}

\bibitem[{{Sloan} {et~al.}(2007){Sloan}, {Jura}, {Duley}, {Kraemer},
  {Bernard-Salas}, {Forrest}, {Sargent}, {Li}, {Barry}, {Bohac}, {Watson}, \&
  {Houck}}]{sjd07}
{Sloan}, G.~C., {Jura}, M., {Duley}, W.~W., {et~al.} 2007, \apj, 664, 1144,
  \dodoi{10.1086/519236}

\bibitem[{{Sloan} {et~al.}(2011){Sloan}, {Hony}, {Smolders}, {Decin},
  {Zijlstra}, {Feast}, {van Wyk}, {van Loon}, {Groenewegen}, \&
  {Sahai}}]{shs11}
{Sloan}, G.~C., {Hony}, S., {Smolders}, K., {et~al.} 2011, \apj, 729, 121,
  \dodoi{10.1088/0004-637X/729/2/121}

\bibitem[{{Sloan} {et~al.}(2014){Sloan}, {Lagadec}, {Zijlstra}, {Kraemer},
  {Weis}, {Matsuura}, {Volk}, {Peeters}, {Duley}, {Cami}, {Bernard-Salas},
  {Kemper}, \& {Sahai}}]{sloanea14}
{Sloan}, G.~C., {Lagadec}, E., {Zijlstra}, A.~A., {et~al.} 2014, \apj, 791, 28,
  \dodoi{10.1088/0004-637X/791/1/28}

\bibitem[{{Sloan} {et~al.}(2016){Sloan}, {Kraemer}, {McDonald}, {Groenewegen},
  {Wood}, {Zijlstra}, {Lagadec}, {Boyer}, {Kemper}, {Matsuura}, {Sahai},
  {Sargent}, {Srinivasan}, {van Loon}, \& {Volk}}]{slo16}
{Sloan}, G.~C., {Kraemer}, K.~E., {McDonald}, I., {et~al.} 2016, \apj, 826, 44,
  \dodoi{10.3847/0004-637X/826/1/44}

\bibitem[{{Smolders} {et~al.}(2010){Smolders}, {Acke}, {Verhoelst},
  {Blommaert}, {Decin}, {Hony}, {Sloan}, {Neyskens}, {van Eck}, {Zijlstra}, \&
  {van Winckel}}]{smoldersea10}
{Smolders}, K., {Acke}, B., {Verhoelst}, T., {et~al.} 2010, \aap, 514, L1,
  \dodoi{10.1051/0004-6361/201014254}

\bibitem[{{Smolders} {et~al.}(2012){Smolders}, {Neyskens}, {Blommaert}, {Hony},
  {van Winckel}, {Decin}, {van Eck}, {Sloan}, {Cami}, {Uttenthaler},
  {Degroote}, {Barry}, {Feast}, {Groenewegen}, {Matsuura}, {Menzies}, {Sahai},
  {van Loon}, {Zijlstra}, {Acke}, {Bloemen}, {Cox}, {De Cat}, {Desmet},
  {Exter}, {Ladjal}, {{\O}stensen}, {Saesen}, {van Wyk}, {Verhoelst}, \&
  {Zima}}]{smoldersea12a}
{Smolders}, K., {Neyskens}, P., {Blommaert}, J.~A.~D.~L., {et~al.} 2012, \aap,
  540, A72, \dodoi{10.1051/0004-6361/201118242}

\bibitem[{{Sylvester} {et~al.}(1994){Sylvester}, {Barlow}, \&
  {Skinner}}]{sylvesterea94}
{Sylvester}, R.~J., {Barlow}, M.~J., \& {Skinner}, C.~J. 1994, \mnras, 266, 640

\bibitem[{{Sylvester} {et~al.}(1998){Sylvester}, {Skinner}, \&
  {Barlow}}]{sylvesterea98}
{Sylvester}, R.~J., {Skinner}, C.~J., \& {Barlow}, M.~J. 1998, \mnras, 301,
  1083, \dodoi{10.1046/j.1365-8711.1998.02078.x}

\bibitem[{{Takigawa} {et~al.}(2015){Takigawa}, {Tachibana}, {Nagahara}, \&
  {Ozawa}}]{takigawa+15}
{Takigawa}, A., {Tachibana}, S., {Nagahara}, H., \& {Ozawa}, K. 2015, \apjs,
  218, 2, \dodoi{10.1088/0067-0049/218/1/2}

\bibitem[{{Tielens}(2008)}]{tielens08}
{Tielens}, A.~G.~G.~M. 2008, \araa, 46, 289,
  \dodoi{10.1146/annurev.astro.46.060407.145211}

\bibitem[{{Ulrich}(1973)}]{ulrich73}
{Ulrich}, R.~K. 1973, in Explosive Nucleosynthesis, ed. D.~N. {Schramm} \&
  W.~D. {Arnett}, 139

\bibitem[{{Verhoelst} {et~al.}(2009){Verhoelst}, {van der Zypen}, {Hony},
  {Decin}, {Cami}, \& {Eriksson}}]{verhoelstea09}
{Verhoelst}, T., {van der Zypen}, N., {Hony}, S., {et~al.} 2009, \aap, 498,
  127, \dodoi{10.1051/0004-6361/20079063}

\bibitem[{{Wallerstein} \& {Knapp}(1998)}]{wk98}
{Wallerstein}, G., \& {Knapp}, G.~R. 1998, \araa, 36, 369,
  \dodoi{10.1146/annurev.astro.36.1.369}

\bibitem[{{Wright} {et~al.}(2010){Wright}, {Eisenhardt}, {Mainzer}, {Ressler},
  {Cutri}, {Jarrett}, {Kirkpatrick}, {Padgett}, {McMillan}, {Skrutskie},
  {Stanford}, {Cohen}, {Walker}, {Mather}, {Leisawitz}, {Gautier}, {McLean},
  {Benford}, {Lonsdale}, {Blain}, {Mendez}, {Irace}, {Duval}, {Liu}, {Royer},
  {Heinrichsen}, {Howard}, {Shannon}, {Kendall}, {Walsh}, {Larsen}, {Cardon},
  {Schick}, {Schwalm}, {Abid}, {Fabinsky}, {Naes}, \& {Tsai}}]{wise10}
{Wright}, E.~L., {Eisenhardt}, P.~R.~M., {Mainzer}, A.~K., {et~al.} 2010, \aj,
  140, 1868, \dodoi{10.1088/0004-6256/140/6/1868}

\bibitem[{{Wurm}(1940)}]{wurm40}
{Wurm}, K. 1940, \apj, 91, 103, \dodoi{10.1086/144150}

\bibitem[{Young {et~al.}(2012)Young, Becklin, Marcum, Roellig, Buizer, Herter,
  Gusten, Dunham, Temi, Andersson, Backman, Burgdorf, Caroff, Casey, Davidson,
  Erickson, Gehrz, Harper, Harvey, Helton, Horner, Howard, Klein, Krabbe,
  McLean, Meyer, Miles, Morris, Reach, Rho, Richter, Roeser, Sandell, Sankrit,
  Savage, Smith, Shuping, Vacca, Vaillancourt, Wolf, \& Zinnecker}]{sofia}
Young, E.~T., Becklin, E.~E., Marcum, P.~M., {et~al.} 2012, \apjl, 749, L17.
\newblock \url{http://stacks.iop.org/2041-8205/749/i=2/a=L17}

\bibitem[{{Zubko} {et~al.}(1996){Zubko}, {Mennella}, {Colangeli}, \&
  {Bussoletti}}]{zubko+96}
{Zubko}, V.~G., {Mennella}, V., {Colangeli}, L., \& {Bussoletti}, E. 1996,
  \mnras, 282, 1321, \dodoi{10.1093/mnras/282.4.1321}

\end{thebibliography}
\bibliographystyle{aasjournal}

\end{document}